\documentclass[aps,amsmath,prl,reprint,showpacs,preprintnumbers,twocolumn]{revtex4}
\usepackage{amsfonts}
\usepackage{graphicx}
\usepackage{dcolumn}
 \usepackage{pdfsync}
\usepackage{multirow}
\usepackage{subfigure}
\usepackage{palatino}
\usepackage{graphicx}
\usepackage{epsfig}
\usepackage{ifpdf}
\ifpdf \DeclareGraphicsExtensions{.pdf,.png,.jpg,.mps}
 \else
\DeclareGraphicsExtensions{.eps} \fi
\begin{document}
\title{Efficiency and Its Bounds for a Quantum Einstein Engine at Maximum Power}
\author{H. Yan}
\affiliation{Department of Physics,
Indiana University/IUCF, 2401 Milo B. Sampson Lane, Bloomington, IN
47408, USA}
\author{Hao Guo$^{1,2}$}
\affiliation{$^1$Department of Physics, Southeast University}
\affiliation{$^2$Department of Physics, the University of Hong Kong}
\email{guohao.ph@gmail.com}

\begin{abstract}
We study a quantum thermal engine model for which the heat transfer law is determined by Einstein's theory of radiation. The working substance of the quantum engine is assumed to be a two-level quantum systems of which the constituent particles obey Maxwell-Boltzmann(M.B.), Fermi-Dirac(F.D.) or Bose-einstein(B.E.) distributions respectively at equilibrium. The thermal efficiency and its bounds at maximum power of these models are derived and discussed in the long and short thermal contact time limits. The similarity and difference between these models are discussed. We also compare the efficiency bounds of this quantum thermal engine to those of its classical counterpart.
\end{abstract}
\pacs{05.70.Ln,05.20.-y}
\maketitle

\section{Introduction}
The Carnot engine plays a crucial role in the foundations of thermodynamics. However, it can not be realized in practice since its output power is infinitesimally small due to its reversibility. Real thermal engines can not work as slowly as Carnot engines to preserve equilibrium and must lose energy during the working cycles due to various reasons. This makes the efficiency of real thermal engines below that of an ideal Carnot engine. To optimize the thermal engines in the real world, a lof of ``realistic'' models have been established and studied in the literature\cite{CUR75,Salamon81,Ondrechen83,ESP10,ESP09,BRO05,PHY,YAN12}. One of the most practical problem associated with the optimization of real heat engines is its efficiency at maximum power. This problem was firstly studied by Curzon and Ahlborn in 1975\cite{CUR75}. For Carnot-like heat engines, the authors assumed that the temperature differences between the heat reservoirs and working substance are finite and fixed, thus the two heat transferring processes are not reversible anymore, while the adiabatic expansion and compression processes are still reversible. Under these assumptions, they derived the well-known CA efficiency $\eta_{\textrm{CA}}=1-\sqrt{T_c/T_h}$, where $T_h$ and $T_c$ are the temperatures of the hot and cold heat reservoirs with which the working substance is in contact. Though the CA formula has a good agreement with measured efficiencies of some thermal plants, this model still has some intrinsic drawbacks.
On one hand, it gives neither an exact nor constraint result for the efficiency as pointed out by Ref.\cite{ESP10}. On the other hand, in real world situations the temperature differences between the working medium and heat reservoirs are not constant and the heat transferring process could be governed by some more general physical laws which can incorporate temperature changing during heat transferring processes.

In our previous work on classical engines\cite{YAN12}, we have seen that the heat transfer law plays a crucial role on the efficiency at maximum power problem.
For Carnot engines, the time periods for which the adiabatic expansion and compression processes last are usually negligibly short, while those of the two isothermal heat transferring processes are infinitely long, therefore Carnot engines have zero output power. In real world situations, the isothermal heat transferring processes must last for a finite period of time and obey specific heat transferring laws.
 In Ref.\cite{YAN12}, we studied a thermal engine model for which Newton's cooling law is obeyed during the heat transferring processes, and derived the upper and lower bounds for the efficiency at maximum power in the long and short contact time limits respectively.
By considering the heat transferring processes during which the temperature of the working medium is close to or far from isothermal, and adjusting the ratio between the heat capacities of the heating and cooling
stages, the model can simulate different types of engines including but not limited to Carnot engines.

The studies of classical thermal engines can be generalized to their quantum counterparts. Recently, different models of quantum thermal engines are extensively studied in the literature\cite{QUA07,QUA09,LU12}. The efficiency of a quantum thermal engine at maximum power has also been studied in Ref.\cite{ABE11}, where the quantum thermal engine is based on the model discussed in Ref.\cite{BEN00},
in which the quantum thermal engine is composed of particles confined in a one-dimensional (1D) infinite potential well, and the wall of the well can expand to perform work. The derived efficiency at maximum power is a universal number.

Following the same spirit, we try to generalize our previous work to the quantum world. For simplicity and without losing generality, the working substance in our model of thermal engines is a two-level quantum system\cite{BEN00}, which can be chosen as the lowest two levels of a 1D infinite quantum well or a 1D harmonic oscillator. Unlike the discussion in Ref.\cite{BEN00}, the temperature rather than average energy is used to describe the thermal equilibrium state as in Ref.\cite{QUA07}.
What is important here is that the heat transferring process between the working substance and the heat reservoir is described by Einstein's theory of radiation. This can be thought as the quantum version of the model discussed in Ref.\cite{YAN12} in some way. For simplicity, we denote this kind of quantum thermal engine by ``quantum Einstein engine''. We are interested in the efficiency and its bounds at maximum power.

The organization of the paper is as follows. We first study the quantum thermal engine of which the constituent particles of the working substance obey the M.B.(Maxwell-Boltzmann) distribution. This will shed light on our successive studies on the other two models. For this model, we derive the heat transferring law based on Einstein's theory of radiation, and give the formulas of the heat transfer and entropy production.  As in Ref.\cite{YAN12}, we study the efficiency at maximum power and its bounds in the long and short contact time limits. We also study quantum engines for which F.D.(Fermi-Dirac) and B.E.(Bose-Einstein) distributions are applied.
\section{Quantum Einstein Engine associated with M.B. distribution}
\subsection{General results for heat transfer and entropy production}
\begin{figure}[htb]
 \includegraphics[width=3.5in, angle=0]{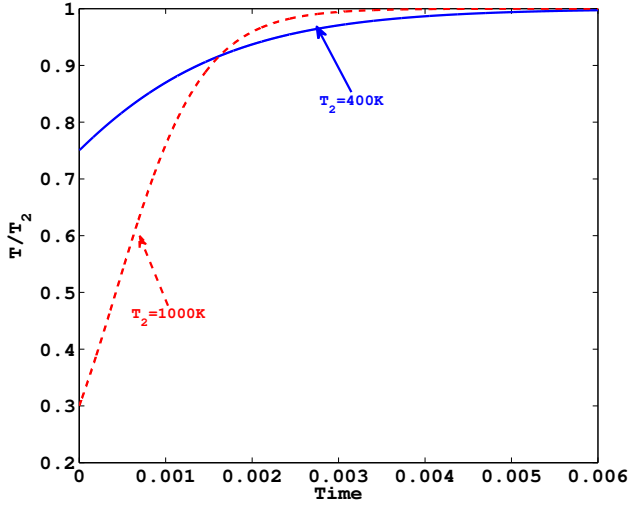}
 \caption{\small{(Color online) Temperature of the working substance during the heat transferring process as a function of time. The blue solid (red dashed) line indicates the temperature changing of the working substance for the situation that the temperature ($T_2$) of the heat reservoir is $400$K ($1000$K). For both situations, the initial temperature of the working substance is $300$K. The unit is chosen such that $A=1$ and $h\nu/k=1$.  }}
 \label{fig.1}
\end{figure}

The working substance of our quantum thermal engine is assumed to be a two-level quantum system with energy levels $E_1$ (low) and $E_2$ (high). The
energy difference of the two levels is $E_{2}-E_{1}=h\nu$, where $h$ is the Planck's constant. The particle numbers at low and high energy levels are $N_{1}$ and $N_{2}$ respectively, and the fixed total particle number is given by $N_{0}=N_{1}+N_{2}$. For simplicity, we first consider the case that the  constituent particles of the quantum system satisfy M.B. distribution at equilibrium states.  Assume the initial temperature of the system is $T_1$, then the initial particle distributions are
\begin{eqnarray}
N_{1}=
N_{0}\frac{1}{1+\exp{(-\beta_{1}h\nu)}}, N_{2}=N_{0}
\frac{1}{1+\exp{(\beta_{1}h\nu)}},
\end{eqnarray}
where $\beta_1=1/k_BT_1$ and $k_B$ is the Boltzmann constant. In our model of quantum thermal engine, the heat reservoir can be thought of as a black-body source with temperature $T_2$.  When the working substance or the two-level quantum system is ``in contact with'' a black-body
source, the heat is transferred by the photon emission and absorbtion. This heat transferring process is described by Einstein's theory of radiation\cite{HEC08}
\begin{eqnarray}\label{E1}
& &\frac{dN_{2}}{dt}=-\frac{dN_{1}}{dt}\nonumber\\
&=&BN_{0}u_{\nu}-2BN_{2}u_{\nu}-AN_{2}=-aN_{2}+b,
\end{eqnarray}
where $a=2Bu_{\nu}+A$, $b=BN_{0}u_{\nu}$. $A$ and $B$ are the famous Einstein's coefficients, and $u_{\nu}$ is the
spectral energy density of the black-body source. The solution of Eq.(\ref{E1}) is given by
\begin{equation}
N_{2}(t)=\frac{b}{a}-[\frac{b}{a}-N_{2}(0)]\exp{(-at)}
\end{equation}
Introducing the distribution function $
f(\beta)=1/(1+\exp{(\beta h\nu)})$ for level $E_2$, then we have
\begin{eqnarray}
N_2(0)=N_0f(\beta_1), \quad N_2(t\rightarrow\infty)=\frac{b}{a}=N_{0}f(\beta_{2}).
\end{eqnarray}
At time $t$, we have $N_2(t)=f(\beta(t))N_0$ where
\begin{eqnarray}\label{E2}
f(\beta(t))=f(\beta_{2})-[f(\beta_{2})-f(\beta_{1})]\exp{(-at)}.
\end{eqnarray}
Similarly, for level $E_1$, we have $N_1(t)=(1-f(\beta(t)))N_0$. We are interested in the situation with $\beta h\nu\ll 1$, which
is true if the size of the quantum well is not too small or the spring constant of the
harmonic oscillator is not too large at high temperature. To the leading order of $\beta h\nu$ we get
\begin{eqnarray}
f(\beta)\approx\frac{1}{2}-\frac{1}{4}\beta h\nu,\quad
a=A\coth{(\beta_{2}h\nu/2)}\approx \frac{2A}{\beta_{2} h\nu}.
\end{eqnarray}
Therefore, Eq.(\ref{E2}) leads to the changing of inverse temperature
\begin{equation}
\beta(t)=\beta_{2}-(\beta_{2}-\beta_{1})\exp{[-\frac{2At}{h\nu\beta_{2}}]}.
\end{equation}
It is interesting to notice that the time constant $1/a$ depends on the temperature of the heat reservoir. If temperature is higher, less time is needed for the working substance to reach equilibrium with the heat reservoir, this phenomenon is  counterintuitive as shown by Figure.\ref{fig.1}.  Note that
in general Einstein's coefficient $A\propto\nu^{2}$\cite{SAL07}, then the inverse time constant $a$ is in fact proportional to $\nu$. Therefore the higher the energy difference $h\nu$, the longer the time is needed for the system to reach the equilibrium state.

In terms of the distribution function $f(\beta)$, the quantum entropy of the two-level system can be expressed as
\begin{equation}\label{E3}
S=-k[f(\beta)\ln{f(\beta)}+(1-f(\beta))\ln{(1-f(\beta))}].
\end{equation}
Plug in the expression of $f(\beta)$, and expand Eq.(\ref{E3}) in series of $\beta h\nu$
\begin{equation}
S=k\ln{2}-\frac{1}{8}k(\beta h\nu)^{2}+\mathcal{O}(\beta h\nu)^{4}.
\end{equation}
For a heat transferring process between the initial state with inverse temperature $\beta_{0}$ and the final state with inverse temperature $\beta(t)$ at time $t$, the heat transfer and entropy production are given by
\begin{eqnarray}
\Delta Q(t)&=&(N_{2}(t)-N_{20})h\nu=N_{0}h\nu(f(\beta(t))-f(\beta_{0}))\nonumber\\
&=&\frac{1}{4}N_{0}h^{2}\nu^{2}(\beta_{0}-\beta(t))+\mathcal{O}\big((\beta h\nu)^{3}\big),\nonumber\\
\Delta S(t)&=&\frac{1}{8}k(h\nu)^{2}(\beta_{0}^{2}-\beta^{2}(t))+\mathcal{O}\big((\beta h\nu)^{4}\big),
\end{eqnarray}
where $N_{20}$ and $N_2(t)$ are the particle numbers of initial and final state respectively. In the derivation of $\Delta Q(t)$, we have used the fact $dN_2=-dN_1$. Generically, the energy levels of the quantum working substance during the heating and cooling stages are different since the engine must expand to perform work between the two stages. We assume that the frequencies associated with the heating and cooling stages are $\nu_h$ and $\nu_c$ respectively, and the initial and final inverse temperatures are $\beta_{h0}$, $\beta_h(t)$ and $\beta_{c0}$, $\beta_c(t)$ respectively.
Therefore, to the leading order of $\beta h\nu$ the heat transfer and entropy production during the heating stage are given by
\begin{eqnarray}
\Delta Q_{h}(t)&=&\frac{1}{4}N_{0}h^{2}\nu_{h}^{2}(\beta_{h0}-\beta_{h}(t)),\nonumber\\
\Delta S_{h}(t)&=&\frac{1}{8}kh^{2}\nu_{h}^{2}(\beta_{h0}^{2}-\beta_{h}(t)^{2}).
\end{eqnarray}
Similarly, during the cooling stage we have
\begin{eqnarray}
\Delta Q_{c}(t)&=&\frac{1}{4}N_{0}h^{2}\nu_{c}^{2}(\beta_{c0}-\beta_{c}(t)),\nonumber\\
\Delta S_{c}(t)&=&\frac{1}{8}kh^{2}\nu_{c}^{2}[\beta_{c0}^{2}-\beta_{c}^{2}(t)].
\end{eqnarray}
We assume the time duration that the heating and cooling stages last are $\tau_h$ and $\tau_c$ respectively. When the quantum engine finishes a full thermodynamical cycle, the working medium returns back to its initial state and we have $\Delta S_h(\tau_h)+\Delta S_c(\tau_c)=0$. The power output and the efficiency of the thermal engine are given by
\begin{eqnarray}\label{E4}
P&=&\frac{\Delta Q_h(\tau_h)+\Delta Q_c(\tau_c)}{\tau_h+\tau_c},\nonumber\\
\eta&=&1+\frac{\Delta Q_c(\tau_c)}{\Delta Q_h(\tau_h)}.
\end{eqnarray}
Here we adopt the convention that $Q>0(<0)$ means absorbing(releasing) heat. As mentioned before, the working substance of the quantum engine under considering is composed by particles at the two lowest levels of a 1D infinite quantum well or 1D harmonic oscillator. The cycles of the quantum engine are as follows. 1)The system absorbs heat $Q_{h}$ from the hot reservoir during the time period $\tau_{h}$. 2)The system expands to perform work. In case of 1D infinite quantum well, the width of the well expands from $L_{0}$ to $L_{1}$. In case of 1D harmonic
oscillator, the``width" of the harmonic potential expands while its spring constant ``shrinks" from $k_{0}$ to $k_{1}$ in its parameter space. 3)The system releases heat $Q_{c}$ at the cold reservoir during the time period $\tau_{c}$.
4)The system returns to its original size and temperature and the working medium returns to its original state.
\subsection{Efficiency and its bounds in the long contact time limit}
When the contact time is long enough, the working substance can exchange heat sufficiently with the
reservoirs. In this limit, $\tau_{h,c}/(\beta h\nu_{h,c} A)$ is large, and we assume the final inverse temperatures for the two stages are $\beta_{h}=\beta_{h}(\tau_{h})$ and $\beta_{c}=\beta_{c}(\tau_{c})$. Hence the heat transfers during the two stages are
\begin{eqnarray}
\Delta Q_{h}=\frac{1}{4}N_{0}h^{2}\nu_{h}^{2}x, \quad
\Delta Q_{c}=-\frac{1}{4}N_{0}h^{2}\nu_{c}^{2}y,
\end{eqnarray}
where $x=\beta_{h0}-\beta_{h}$ and $y=\beta_{c}-\beta_{c0}$.
To the leading order of $\beta h\nu$, the constraint $\Delta S_{h}+\Delta S_{c}=0$ gives
\begin{eqnarray}
\nu_{h}^{2}(2\beta_{h}+x)x-\nu_{c}^{2}(2\beta_{c}-y)y=0,
\end{eqnarray}
of which the solution is given by
\begin{eqnarray}\label{E5}
x=\sqrt{\beta_{h}^{2}+\frac{1}{\gamma^{2}}(2\beta_{c}-y)y}-\beta_{h},
\end{eqnarray}
where $\gamma=\nu_h/\nu_c$. Substitute Eq.(\ref{E5}) to the expression of the output power
\begin{eqnarray}
P=\frac{1}{4}N_{0}h^{2}\frac{\nu_{h}^{2}x-\nu_{c}^{2}y}{\tau_{h}+\tau_{c}}
\end{eqnarray}
and let $\partial{P}/\partial{y}=0$, the only meaningful solution for $y$ is found to be
\begin{equation}
y=\frac{\beta_{c}(1+\gamma^{2})-\sqrt{\beta_{c}^{2}(1+\gamma^{2})+\beta_{h}^{2}\gamma^{2}(1+\gamma^{2})}}{1+\gamma^{2}}
\end{equation}
Plug $y$ into Eq.(\ref{E4}) we obtain the thermal efficiency at maximum power
\begin{eqnarray}\label{E9}
\eta_{m}&=&1-\frac{1}{\gamma^{2}}\frac{y}{x}\\
&=&1-\frac{\gamma^{2}(1-\eta_{c})-1+\sqrt{(1+\gamma^{2})[1+\gamma^{2}(1-\eta_{c})^{2}]}}{(2-\eta_{c})\gamma^{2}},\nonumber
\end{eqnarray}
where $\eta_c=1-\frac{T_c}{T_h}$ is the Carnot efficiency. If we let $\gamma$ approach $0$ and $\infty$ respectively, we obtain the upper and lower bounds of $\eta_m$ as
\begin{equation}
\frac{\eta_{c}}{2}\leq\eta_{m}\leq\frac{\eta_{c}}{2-\eta_{c}}.
\end{equation}
Interestingly, these bounds agree exactly with those given in Ref.\cite{ESP10} for classical thermal engines in the long contact time limit. However, the situation is a little different for our model of quantum thermal engine.
 The quantum engine, either quantum-well type or harmonic-oscillator type, must expand to perform work after absorbing heat at hot reservoir. Hence its size must increase afterwards. For a quantum-well type engine, its width $L$ will increase,
while for a harmonic-oscillator type engine, its spring constant $k$ will decrease.
Since the frequency is anti-proportional to $L^2$ or proportional to $\sqrt{k}$, then $\gamma=\nu_{h}/\nu_{c}$ must be larger than $1$, or the lower limit of $\gamma$ is $1$ rather than $0$. Now we have tighter bounds for $\eta_m$
\begin{equation}
\frac{2-\sqrt{4-4\eta_{c}+2\eta_{c}^{2}}}{2-\eta_{c}}\leq\eta_{m}\leq\frac{\eta_{c}}{2-\eta_{c}}.
\end{equation}
The efficiency $\eta_{m}$ can be expanded in series of $\eta_{c}$ as
\begin{equation}
\eta_{m}=\frac{1}{2}\eta_{c}+\frac{\gamma^{2}}{4(1+\gamma^{2})}\eta_{c}^{2}+\mathcal{O}(\eta_{c}^{3}).
\end{equation}
The coefficient of the second-order term lies between $1/8$ and $1/4$.
\subsection{Efficiency and its bounds in the short contact time limit}
In the short time limit such that $\tau\ll 1/a$, the inverse temperature can be approximated to the second order of $a\tau$ as
\begin{eqnarray}
\beta(\tau)
&\approx&\beta_{1}+(\beta_{2}-\beta_{1})a\tau+(\beta_{1}-\beta_{2})\frac{1}{2}a^{2}\tau^{2}.
\end{eqnarray}
Implementing this approximation, the entropy productions during the two stages are given by
\begin{eqnarray}
\Delta S_{h}&=&\nu_{h}^{2}[-2\beta_{h0}(\beta_{h}-\beta_{h0})a_{h}\tau_{h}\nonumber\\
& &-(\beta_{h0}-\beta_{h})(2\beta_{h0}-\beta_{h})a_{h}^{2}\tau_{h}^{2}],\nonumber\\
\Delta S_{c}&=&\nu_{c}^{2}[-2\beta_{c0}(\beta_{c}-\beta_{c0})a_{c}\tau_{c}\nonumber\\
& &-(\beta_{c0}-\beta_{c})(2\beta_{c0}-\beta_{c})a_{c}^{2}\tau_{c}^{2}].
\end{eqnarray}
Using the same convention as in the last subsection, the constraint $\Delta S_{h}+\Delta S_{c}=0$ gives an equation for $x$ and $y$. Since $a_{h,c}\tau_{h,c}$ is an infinitesimal quantity, we match both sides of the equation order by order of $a_{h,c}\tau_{h,c}$. To the first and second order, we get
\begin{eqnarray}
\gamma^{2}(\beta_{h}+x)x\frac{A_{h}}{\beta_{h}\nu_{h}}\tau_{h}&=&(\beta_{c}-y)y\frac{A_{c}}{\beta_{c}\nu_{c}}\tau_{c},\nonumber\\
\gamma^{2}x(\beta_{h}+2x)a_{h}^{2}\tau_{h}^{2}&=&(\beta_{c}-2y)ya_{c}^{2}\tau_{c}^{2},
\end{eqnarray}
from which one can deduce
\begin{equation}\label{E6}
\frac{\beta_{h}+2x}{x(\beta_{h}+x)^{2}}=\gamma^{2}\frac{\beta_{c}-2y}{y(\beta_{c}-y)^{2}}.
\end{equation}
The key step here is to simplify the above equation and get a relatively simple relation between $x$ and $y$ as in Ref.\cite{YAN12}, hence we can avoid messing up the physics by the mathematical complexity. We assume that the temperature difference is small relative
to the temperature of the heat reservoir at each heat transferring stage. Thus, $x$ is small relative to $\beta_{h}$ and $y$ small to $\beta_{c}$. Expanding both sides of Eq.(\ref{E6}) to the third order of $x$ and $y$, we can derive a very simple relation between $x$ and $y$
\begin{equation}
\frac{y}{x}=\gamma^{2}\frac{\beta_{h}}{\beta_{c}}.
\end{equation}
The heat transfers during the two heating and cooling stages are given by
\begin{eqnarray}
\Delta Q_{h}=\frac{1}{4}N_{0}h^{2}\nu_{h}^{2}xa_{h}\tau_{h},\quad
\Delta Q_{c}=-\frac{1}{4}N_{0}h^{2}\nu_{c}^{2}ya_{c}\tau_{c},
\end{eqnarray}
from which the output power is given by
\begin{equation}
P=\frac{1}{4}N_{0}h^{2}\frac{\nu_{h}^{2}xa_{h}\tau_{h}-\nu_{c}^{2}ya_{c}\tau_{c}}{\tau_{h}+\tau_{c}}.
\end{equation}
Taking into account that the spontaneous emission coefficient $A$ satisfies $
A\propto\nu^{2}$,
except a constant the output power is evaluated as
\begin{eqnarray}
P\propto\frac{(\beta_{c}-\gamma^{2}x\beta_{h}/\beta_{c}-\beta_{h}-x)x}{\beta_{c}-\gamma^{2}x\beta_{h}/\beta_{c}+\gamma(\beta_{h}+x)\beta_{c}^{2}/\beta_{h}^{2}}.
\end{eqnarray}
Solve the equation $\partial P/\partial x=0$, and plug the only meaningful solution for $x$ into the expression of the efficiency at maximum power
\begin{equation}
\eta_{m}=1-\frac{\beta_{h}+x}{\beta_{c}-\gamma^{2}x\beta_{h}/\beta_{c}}.
\end{equation}
In series of $\eta_{c}$, $\eta_{m}$ can be expanded as
\begin{equation}\label{E10}
\eta_{m}=\frac{\eta_{c}}{2}+\frac{\gamma(2\gamma^{2}+\gamma+1)}{8(1+\gamma+\gamma^{2}+\gamma^{3})}\eta_{c}^{2}+\mathcal{O}(\eta_{c}^3).
\end{equation}
Taking $\gamma=0$, $\infty$, the bounds of $\eta_m$ are derived as
\begin{equation}
\frac{\eta_{c}}{2}\leq\eta_{m}\leq\frac{\eta_{c}}{2-\eta_{c}}.
\end{equation}
Interestingly, we obtain the same rough bounds of $\eta_m$ as in the long time contact limit. Again noting that $\gamma$ can not be smaller than $1$, the finer bounds of $\eta_m$ are found to be
\begin{equation}
1-\sqrt{1-\eta_{c}}\leq\eta_{m}\leq\frac{\eta_{c}}{2-\eta_{c}}.
\end{equation}
The coefficient of the second-order term of $\eta_{m}$ also lies between $1/8$ and $1/4$.
\section{Quantum Einstein Engine associated with the F.D. distribution}
In this section, we consider the case that the constituent particles of the working substance obey the F.D distribution. In fact we will see that this situation will reduce to that associated with the M.B. distribution even when $\beta E_{1,2}\ll 1$ (If $\beta E_{1,2}\gg 1$, both F.D. and B.E. distributions reduce to M.B. distribution).
Here we still adopt the same convention for the parameters as in the last section. For each level of the quantum system, the particle numbers of the initial state are given by
\begin{eqnarray}
N_1(0)=\frac{N_0}{1+\frac{e^{\beta_1E_1}+1}{e^{\beta_1E_2}+1}},\quad N_2(0)=\frac{N_0}{1+\frac{e^{\beta_1E_2}+1}{e^{\beta_1E_1}+1}},
\end{eqnarray}
Similar to what we have done in the last section, the distribution function $f^{\textrm{F}}(\beta)=1/(1+\frac{e^{\beta E_2}+1}{e^{\beta E_1}+1})$ is introduced. Thus the initial number distributions are $N_1(0)=N_0(1-f^{\textrm{F}}(\beta_1))$ and $N_2(0)=N_0f^{\textrm{F}}(\beta_1)$. When the quantum system is in contact with the hot (cold) reservoir, the heating (cooling) process is described by Einstein's theory of radiation Eq.(\ref{E1}). Solve this equation, we get $N_1(t)=N_0(1-f^{\textrm{F}}(\beta(t)))$ and $N_2(t)=N_0f^{\textrm{F}}(\beta(t))$ in which $f^{\textrm{F}}(\beta(t))$ is also expressed by Eq.(\ref{E2}) with $\beta(0)=\beta_1$ and $\beta(\infty)=\beta_2$. Hence the heat absorbed by the working substance at time $t$ is
$\Delta Q(t)=N_0h \nu \big(f^{\textrm{F}}(\beta(t))-f^{\textrm{F}}(\beta(0))\big)$. To the leading order of $\beta h\nu$ we have
\begin{eqnarray}
f^{\textrm{F}}(\beta)\approx\frac{1}{2}-\frac{1}{8}\beta h\nu, \\
\quad a^{\textrm{F}}=\frac{e^{\beta_2E_2}+e^{\beta_2E_1}+2}{e^{\beta_2E_2}-e^{\beta_2E_1}}\approx \frac{4A}{\beta_2 h\nu}.
\end{eqnarray}
The quantum entropy of the working substance is approximated by
\begin{eqnarray}
S^{\textrm{F}}\approx \ln2-\frac{1}{32}\beta^2h^2\nu^2.
\end{eqnarray}
One can see that the expressions of $f^{\textrm{F}}(\beta)$, $a^{\textrm{F}}$ and $S^{\textrm{F}}$ differ from their maxwellian counterparts only in the coefficients of the leading order of $\beta h\nu$. If we go on carrying the calculations as before, it is not difficult to find that the results are the same as those associated with M.B. distribution. In other word, the quantum Einstein engine with fermionic working substance has no significant difference from that with the maxwellian working substance if $\beta E_{1,2}\ll 1$.
\section{Quantum Einstein Engine associated with the B.E. distribution}
\subsection{General results for heat transfer and entropy production}
Now we consider the the quantum Einstein engine of which the working substance obeys the B.E. distribution. The discussions follow quite similar steps as the previous section. The initial particle distributions of the two-level quantum systems are
\begin{eqnarray}
N_1(0)=\frac{N_0}{1+\frac{e^{\beta_1E_1}-1}{e^{\beta_1E_2}-1}},\quad N_2(0)=\frac{N_0}{1+\frac{e^{\beta_1E_2}-1}{e^{\beta_1E_1}-1}},
\end{eqnarray}
The heating (cooling) process is again governed by Einstein's theory of radiation. Introducing the distribution function $f^{\textrm{B}}(\beta)=1/(1+\frac{e^{\beta E_2}-1}{e^{\beta E_1}-1})$, then at time $t$ the particle distributions become $N_1(t)=N_0(1-f^{\textrm{B}}(\beta(t)))$ and $N_2(t)=N_0f^{\textrm{B}}(\beta(t))$ with $\beta(0)=\beta_1$ and $\beta(\infty)=\beta_2$. The heat transfer is given by $\Delta Q(t)=N_0h \nu \big(f^{\textrm{B}}(\beta(t))-f^{\textrm{B}}(\beta(0))\big)$.
Similarly, to the leading order of $\beta h\nu$ we have
\begin{eqnarray}
f^{\textrm{B}}(\beta)&\approx&\frac{E_1}{E_1+E_2}-\frac{1}{2}\frac{h\nu\beta E_1E_2}{(E_1+E_2)^2}, \nonumber\\
a^{\textrm{B}}=\frac{e^{\beta_2E_2}+e^{\beta_2E_1}-2}{e^{\beta_2E_2}-e^{\beta_2E_1}}&\approx& \frac{A(E_1+E_2)}{h\nu}.
\end{eqnarray}
The quantum entropy production is further approximated by
\begin{small}
\begin{widetext}
\begin{equation}
S^{\textrm{B}}\approx-\Big[\frac{E_1}{E_1+E_2}\ln\frac{E_1}{E_1+E_2}+\frac{E_2}{E_1+E_2}\ln\frac{E_2}{E_1+E_2}\Big]-\frac{E_1E_2\ln\frac{E_2}{E_1}h\nu}{2(E_1+E_2)^2}\beta
-\frac{E_1E_2h\nu\big(3(E^2_2-E^2_1)+2(E^2_1-4E_1E_2+E^2_2)\ln\frac{E_1}{E_2}\big)}{24(E_1+E_2)^3}\beta^2.
\end{equation}
\end{widetext}
\end{small}
In what follows, we will ignore the superscript ``B'' for simplicity. For the heating stage, we use $E_{h0}$ and $E_h$ to denote the initial (low) and final (high) energy levels, $\beta_{h0}$ and $\beta_h(t)$ to denote the initial and final inverse temperatures. Define the frequency $\nu_h$ by $E_h-E_{h0}=h\nu_h$, then the heat transfer and entropy production during the heating  and cooling stages are given by
\begin{eqnarray}
\Delta Q_h(t)&=&\frac{1}{2}N_0h^2\nu^2_hX_h\big(\beta_{h0}-\beta_h(t)\big),\nonumber\\
\Delta S_h(t)&=&Y_{h1}\big(\beta_{h0}-\beta_h(t)\big)+Y_{h2}\big(\beta^2_{h0}-\beta^2_h(t)\big),\\
\Delta Q_c(t)&=&\frac{1}{2}N_0h^2\nu^2_cX_c\big(\beta_{c0}-\beta_c(t)\big),\\
\Delta S_c(t)&=&Y_{c1}\big(\beta_{c0}-\beta_c(t)\big)+Y_{c2}\big(\beta^2_{c0}-\beta^2_c(t)\big),
\end{eqnarray}
where
\begin{footnotesize}
\begin{eqnarray*}
X_h&=&\frac{E_{h0}E_h}{(E_{h0}+E_h)^2},\quad Y_{h1}=\frac{E_hE_{h0}\ln\frac{E_h}{E_{h0}}h\nu_h}{2(E_h+E_{h0})^2},\nonumber\\
Y_{h2}&=&\frac{E_{h0}E_hh\nu_h\big(3(E^2_h-E^2_{h0})+2(E^2_{h0}-4E_{h0}E_h+E^2_h)\ln\frac{E_{h0}}{E_h}\big)}{24(E_{h0}+E_h)^3},\\
X_c&=&\frac{E_{c0}E_c}{(E_{c0}+E_c)^2},\quad Y_{c1}=\frac{E_cE_{c0}\ln\frac{E_c}{E_{c0}}h\nu_c}{2(E_c+E_{c0})^2},\nonumber\\
Y_{c2}&=&\frac{E_{c0}E_ch\nu_c\big(3(E^2_c-E^2_{c0})+2(E^2_{c0}-4E_{c0}E_h+E^2_c)\ln\frac{E_{c0}}{E_c}\big)}{24(E_{c0}+E_c)^3}.
\end{eqnarray*}
\end{footnotesize}

\subsection{Efficiency and its bounds in the long contact time limit}
From now on, the discussions are simply parallel to what we have done in the last section. We briefly outline our results here. In this limit, $\tau_{h,c}\rightarrow \infty$. As previously did, we assume $\beta_h(\tau_h)=\beta_h$, $\beta_c(\tau_c)=\beta_c$, $\beta_{h0}-\beta_h=x$ and $\beta_{c0}-\beta_c=-y$.
From $\Delta S_h+\Delta S_c=0$ we have
\begin{eqnarray}
Y_{h1}x+Y_{h2}(2\beta_h+x)x-Y_{c1}y-Y_{c2}(2\beta_c-y)y=0.
\end{eqnarray}
Solve this equation, we have
\begin{eqnarray}\label{E7}
x=\sqrt{\beta^{\prime2}_h+\frac{1}{\gamma_1^2}(2\beta^{\prime}_c-y)y}-\beta'_h,
\end{eqnarray}
\begin{eqnarray}
\beta'_h=\beta_h+\frac{Y_{h1}}{2Y_{h2}}, \beta'_c=\beta_c+\frac{Y_{c1}}{2Y_{c2}}, \gamma^2_1=\frac{Y_{h2}}{Y_{c2}}.
\end{eqnarray}
The power of this heat engine is given by
\begin{eqnarray}
P=\frac{1}{2}N_0h^2\frac{X_h\nu^2_hx-X_c\nu^2_cy}{\tau_h+\tau_c}.
\end{eqnarray}
It is maximized when $\partial P/\partial y=0$. Let $\gamma^2_2=\frac{X_h}{X_c}\frac{\nu^2_h}{\nu^2_c}$, then we have $\frac{\partial x}{\partial y}-\frac{1}{\gamma^2_2}=0$. Using Eq.(\ref{E7}), the only meaningful solution for $y$ is
\begin{eqnarray}\label{y}
y=\frac{\beta'_c(\gamma^2_1+\gamma^4_2)-\sqrt{(\beta^{\prime2}_c\gamma^2_1+\beta^{\prime2}_h\gamma^4_1)(\gamma^2_1+\gamma^4_2})}{\gamma^2_1+\gamma^4_2}.
\end{eqnarray}
The efficiency of the thermal engine at maxim power is
\begin{equation}
\eta_m=1-\frac{X_c\nu^2_c}{X_h\nu^2_h}\frac{y}{x}=1-\frac{1}{\gamma^2_2}\frac{y}{x}.
\end{equation}
Plugging Eqs.(\ref{E7}) and (\ref{y}) we have
\begin{small}
\begin{equation}\label{E8}
\eta_m=1-\frac{\gamma^2_1}{\gamma^2_2}\frac{\gamma^2_2(1-\eta'_c)-1+\frac{1}{\gamma_1}\sqrt{\big(1+\gamma^2_1(1-\eta'_c)^2\big)(\gamma^2_1+\gamma^4_2)}}{\gamma^2_2+\gamma^2_1(1-\eta'_c)},
\end{equation}
\end{small}
where $\eta'_c=1-\frac{\beta'_h}{\beta'_c}$ can be thought of as the corrected Carnot efficiency. If we chose suitable $E_{h,c}$ and $E_{h0, c0}$ such that $\gamma_1=\gamma_2=\gamma'$, Eq.(\ref{E8}) ``recovers'' the result (\ref{E9}) except that $\eta_c$ is replaced by $\eta'_c$ and $\gamma$ by $\gamma'$. By inspecting the numerators and denominators of $\gamma_1$ and $\gamma_2$, one can find that they are of the same order of $E_{h,c}$ and $E_{h0,c0}$ respectively. Hence it is reasonable to assume that they are of the same order when approaching $0$ or $\infty$. Therefore we get a rough estimation of the upper and lower bounds
\begin{equation}
\frac{\eta'_c}{2}\le\eta_m\le\frac{\eta'_c}{2-\eta'_c}.
\end{equation}
Following the same reasoning as before, the tighter bounds of $\eta_m$ are given by
\begin{equation}
\frac{2-\sqrt{4-4\eta_{c}'+2\eta_{c}^{\prime2}}}{2-\eta'_{c}}\leq\eta_{m}\leq\frac{\eta'_{c}}{2-\eta'_{c}}.
\end{equation}
Despite the similarity between the results associated with the B.E. distribution and those associated with the M.B. and F.D. distributions, there is also qualitative difference between them. Obviously the latter only depend on the difference of energy levels, or $\nu_{h,c}$. However, the former has an explicit dependence on the choice of initial (low) energy level $E_{h0,c0}$.

\subsection{Efficiency and its bounds in the short contact time limit}
We consider the limit that $\tau\ll 1/a$ and expand the inverse temperature to the second order of $a\tau$, then the entropy productions during the two stages are given by
\begin{eqnarray}
\Delta S_{h}&=&-2Y_{h2}(\beta'_{h}+x)xa_{h}\tau_{h}-Y_{h2}(\beta'_h+2x)xa_{h}^{2}\tau_{h}^{2},\nonumber\\
\Delta S_{c}&=&2Y_{c2}(\beta'_{c}-y)ya_{c}\tau_{c}+Y_{c2}(\beta'_c-2y)ya_{c}^{2}\tau_{c}^{2}.
\end{eqnarray}
While the heat transfers are
\begin{eqnarray}
\Delta Q_{h}&=&\frac{1}{2}N_{0}h^{2}\nu_{h}^{2}X_hxa_{h}\tau_{h},\nonumber\\
\Delta Q_{c}&=&-\frac{1}{2}N_{0}h^{2}\nu_{c}^{2}X_cya_{c}\tau_{c}.
\end{eqnarray}
As before we match both sides of $\Delta S_h+\Delta S_c=0$ order by order of $a\tau$, then the first and second orders of $a\tau$ give
\begin{eqnarray}
\gamma^2_1(\beta'_{h}+x)xa_{h}\tau_{h}&=&(\beta'_{c}-y)ya_{c}\tau_{c},\nonumber\\
\gamma^2_1(\beta'_{h}+2x)xa^2_{h}\tau^2_{h}&=&(\beta'_{c}-2y)ya^2_{c}\tau^2_{c},
\end{eqnarray}
from which we deduce
\begin{equation}
\frac{\beta'_{h}+2x}{x(\beta'_{h}+x)^{2}}=\gamma^{2}_1\frac{\beta'_{c}-2y}{y(\beta'_{c}-y)^{2}}.
\end{equation}
Using the same argument as in the last section, one can find
\begin{equation}
\frac{y}{x}=\gamma^{2}_1\frac{\beta'_{h}}{\beta'_{c}}.
\end{equation}
The spontaneous emission coefficient $A$ satisfies $A\propto \nu^2$, then the output power is evaluated as:
\begin{small}
\begin{equation}
P=\frac{1}{2}N_0h^2\nu^2_hX_ha_h\frac{\big(\beta'_{c}-\gamma^{2}_1x\beta'_{h}/\beta'_{c}-\frac{\gamma^2_1}{\gamma^2_2}(\beta'_{h}+x)\big)x}{\beta'_{c}-\gamma^{2}_1x\beta'_{h}/\beta'_{c}+\gamma_3(\beta'_{h}+x)\beta'_{c}/\beta'_{h}},
\end{equation}
\end{small}
where $\gamma_3=\frac{\nu_h(E_{h0}+E_h)}{\nu_c(E_{c0}+E_c)}$. The power is maximized when $\partial P/\partial x=0$, which leads to the solution
of the efficiency at maximum power is
\begin{small}
\begin{eqnarray}
\eta_m=\frac{\beta'_h\gamma^2_1+\beta'_c\gamma^2_2\gamma_3-\gamma_1\sqrt{\frac{\beta^{\prime}_c\beta'_h(\beta'_c+\beta'_h\gamma^2_2)(1+\gamma_3)(\beta'_h\gamma^2_1+\beta'_c\gamma^2_2\gamma_3)}{\beta^{\prime2}_c+\beta^{\prime2}_h\gamma^2_1}}}{\gamma^2_2\big[\beta'_c\gamma_3-\beta'_h\gamma_1\sqrt{\frac{\beta^{\prime}_c\beta'_h(1+\gamma_3)(\beta'_h\gamma^2_1+\beta'_c\gamma^2_2\gamma_3)}{(\beta^{\prime2}_c+\beta^{\prime2}_h\gamma^2_1)(\beta'_c+\beta'_h\gamma^2_2)}}\big]}.
\end{eqnarray}
\end{small}
If $E_{h,c}$ and $E_{h0,c0}$ are chosen suitably such that $\gamma_1=\gamma_2=\gamma_3=\gamma'$, the efficiency $\eta_m$ can also be expanded in series of $\eta'_c$ as
\begin{eqnarray}
\eta_m=\frac{\eta'_c}{2}+\frac{\gamma'(2\gamma^{\prime2}+\gamma'+1)}{8(1+\gamma'+\gamma^{\prime2}+\gamma^{\prime3})}\eta^{\prime2}_c+\mathcal{O}(\eta^{\prime3}_c).
\end{eqnarray}
Interestingly, this also ``recovers'' the result (\ref{E10}) expect that $\eta_c$ is replaced by $\eta'_c$ and $\gamma$ by $\gamma'$. Similarly, when $\gamma'$ approaches $0$, $\infty$ or $1$, $\infty$, the rough and fine bounds of $\eta_m$ are found to be
\begin{eqnarray}
\frac{\eta'_c}{2}\leq&\eta_m&\le\frac{\eta'_c}{2-\eta'_c},\nonumber\\
1-\sqrt{1-\eta'_{c}}\leq&\eta_{m}&\leq\frac{\eta'_{c}}{2-\eta'_{c}}.
\end{eqnarray}
\section{Conclusions}
In conclusion, we presented an analysis of the quantum thermal engine of which the heat transferring law is derived from Einstein's theory of radiation. Again we notice that heat transferring laws play a crucial role on the problems about the efficiency of quantum thermal engines at maximum power.The thermal efficiency and its bounds at maximum power for quantum Einstein engines are studied in the long and short time limits.
To some extent, this can be thought as the quantum counterpart of the classical thermal engine studied in Ref.\cite{YAN12}, which can simulate some well-known classical thermal engines. For $\beta h\nu\ll1$,
We find that the quantum Einstein engine with fermionic working substance has no difference from that with maxwellian working substance, while the one with bosonic working substance has a qualitative difference.\\
H. Yan thanks the support by U.S. Department of Energy, Office of Science under grant DE-FG02-03ER46093. Hao Guo thanks the support by Natural Science Foundation of Jiangsu Province, China (SBK201241926). H. Yan thanks professor M.W. Snow for support.

\bibliographystyle{apsrev}

\begin{thebibliography}{15}
\expandafter\ifx\csname natexlab\endcsname\relax\def\natexlab#1{#1}\fi
\expandafter\ifx\csname bibnamefont\endcsname\relax
  \def\bibnamefont#1{#1}\fi
\expandafter\ifx\csname bibfnamefont\endcsname\relax
  \def\bibfnamefont#1{#1}\fi
\expandafter\ifx\csname citenamefont\endcsname\relax
  \def\citenamefont#1{#1}\fi
\expandafter\ifx\csname url\endcsname\relax
  \def\url#1{\texttt{#1}}\fi
\expandafter\ifx\csname urlprefix\endcsname\relax\def\urlprefix{URL }\fi
\providecommand{\bibinfo}[2]{#2}
\providecommand{\eprint}[2][]{\url{#2}}

\bibitem[{\citenamefont{Curzon and Ahlborn}(1975)}]{CUR75}
\bibinfo{author}{\bibfnamefont{F.}~\bibnamefont{Curzon}} \bibnamefont{and}
  \bibinfo{author}{\bibfnamefont{B.}~\bibnamefont{Ahlborn}},
  \bibinfo{journal}{Am. J. Phys.} \textbf{\bibinfo{volume}{43}},
  \bibinfo{pages}{22} (\bibinfo{year}{1975}).

\bibitem[{\citenamefont{Salamon and Nitzan}(1981)}]{Salamon81}
\bibinfo{author}{\bibfnamefont{P.}~\bibnamefont{Salamon}} \bibnamefont{and}
  \bibinfo{author}{\bibfnamefont{A.}~\bibnamefont{Nitzan}},
  \bibinfo{journal}{J. Chem. Phys.} \textbf{\bibinfo{volume}{74}},
  \bibinfo{pages}{3546} (\bibinfo{year}{1981}).

\bibitem[{\citenamefont{Ondrechen et~al.}(1983)\citenamefont{Ondrechen, Rubin,
  and Band}}]{Ondrechen83}
\bibinfo{author}{\bibfnamefont{M.~J.} \bibnamefont{Ondrechen}},
  \bibinfo{author}{\bibfnamefont{M.~H.} \bibnamefont{Rubin}}, \bibnamefont{and}
  \bibinfo{author}{\bibfnamefont{Y.~B.} \bibnamefont{Band}},
  \bibinfo{journal}{J. Chem. Phys.} \textbf{\bibinfo{volume}{78}},
  \bibinfo{pages}{4721} (\bibinfo{year}{1983}).

\bibitem[{\citenamefont{Esposito et~al.}(2010)\citenamefont{Esposito, Kawai,
  Lindenberg, and Van~den Broeck}}]{ESP10}
\bibinfo{author}{\bibfnamefont{M.}~\bibnamefont{Esposito}},
  \bibinfo{author}{\bibfnamefont{R.}~\bibnamefont{Kawai}},
  \bibinfo{author}{\bibfnamefont{K.}~\bibnamefont{Lindenberg}},
  \bibnamefont{and} \bibinfo{author}{\bibfnamefont{C.}~\bibnamefont{Van~den
  Broeck}}, \bibinfo{journal}{Phys. Rev. Lett.} \textbf{\bibinfo{volume}{105}},
  \bibinfo{pages}{150603} (\bibinfo{year}{2010}).

\bibitem[{\citenamefont{Esposito et~al.}(2009)\citenamefont{Esposito,
  Lindenberg, and VandenBroeck}}]{ESP09}
\bibinfo{author}{\bibfnamefont{M.}~\bibnamefont{Esposito}},
  \bibinfo{author}{\bibfnamefont{K.}~\bibnamefont{Lindenberg}},
  \bibnamefont{and}
  \bibinfo{author}{\bibfnamefont{C.}~\bibnamefont{VandenBroeck}},
  \bibinfo{journal}{Phys. Rev. Lett.} \textbf{\bibinfo{volume}{102}},
  \bibinfo{pages}{130602} (\bibinfo{year}{2009}).

\bibitem[{\citenamefont{VandenBroeck}(2005)}]{BRO05}
\bibinfo{author}{\bibfnamefont{C.}~\bibnamefont{VandenBroeck}},
  \bibinfo{journal}{Phys. Rev. Lett.} \textbf{\bibinfo{volume}{95}},
  \bibinfo{pages}{190602} (\bibinfo{year}{2005}).

\bibitem[{PHY(2010)}]{PHY}
\bibinfo{journal}{Phys. Today} \textbf{\bibinfo{volume}{63}},
  \bibinfo{pages}{11} (\bibinfo{year}{2010}).

\bibitem[{\citenamefont{Yan and Guo}(2012)}]{YAN12}
\bibinfo{author}{\bibfnamefont{H.}~\bibnamefont{Yan}} \bibnamefont{and}
  \bibinfo{author}{\bibfnamefont{H.}~\bibnamefont{Guo}},
  \bibinfo{journal}{Phys. Rev. E} \textbf{\bibinfo{volume}{85}},
  \bibinfo{pages}{011146} (\bibinfo{year}{2012}).

\bibitem[{\citenamefont{Quan et~al.}(2007)\citenamefont{Quan, Liu, Sun, and
  Franco}}]{QUA07}
\bibinfo{author}{\bibfnamefont{H.~T.} \bibnamefont{Quan}},
  \bibinfo{author}{\bibfnamefont{Y.}~\bibnamefont{Liu}},
  \bibinfo{author}{\bibfnamefont{C.~P.} \bibnamefont{Sun}}, \bibnamefont{and}
  \bibinfo{author}{\bibfnamefont{N.}~\bibnamefont{Franco}},
  \bibinfo{journal}{Phys. Rev. E} \textbf{\bibinfo{volume}{76}},
  \bibinfo{pages}{031105} (\bibinfo{year}{2007}).

\bibitem[{\citenamefont{Quan}(2009)}]{QUA09}
\bibinfo{author}{\bibfnamefont{H.~T.} \bibnamefont{Quan}},
  \bibinfo{journal}{Phys. Rev. E} \textbf{\bibinfo{volume}{79}},
  \bibinfo{pages}{041129} (\bibinfo{year}{2009}).

\bibitem[{\citenamefont{Lu and Long}(2012)}]{LU12}
\bibinfo{author}{\bibfnamefont{Y.}~\bibnamefont{Lu}} \bibnamefont{and}
  \bibinfo{author}{\bibfnamefont{G.~L.} \bibnamefont{Long}},
  \bibinfo{journal}{Phys. Rev. E} \textbf{\bibinfo{volume}{85}},
  \bibinfo{pages}{011125} (\bibinfo{year}{2012}).

\bibitem[{\citenamefont{Abe}(2011)}]{ABE11}
\bibinfo{author}{\bibfnamefont{S.}~\bibnamefont{Abe}}, \bibinfo{journal}{Phys.
  Rev. E} \textbf{\bibinfo{volume}{83}}, \bibinfo{pages}{041117}
  (\bibinfo{year}{2011}).

\bibitem[{\citenamefont{Bender et~al.}(2000)\citenamefont{Bender, Brody, and
  Meister}}]{BEN00}
\bibinfo{author}{\bibfnamefont{C.~M.} \bibnamefont{Bender}},
  \bibinfo{author}{\bibfnamefont{D.~C.} \bibnamefont{Brody}}, \bibnamefont{and}
  \bibinfo{author}{\bibfnamefont{B.~K.} \bibnamefont{Meister}},
  \bibinfo{journal}{J. Phys. A} \textbf{\bibinfo{volume}{33}},
  \bibinfo{pages}{4427} (\bibinfo{year}{2000}).

\bibitem[{\citenamefont{Hecht and Ganesan}(2008)}]{HEC08}
\bibinfo{author}{\bibfnamefont{E.}~\bibnamefont{Hecht}} \bibnamefont{and}
  \bibinfo{author}{\bibfnamefont{A.~R.} \bibnamefont{Ganesan}},
  \emph{\bibinfo{title}{Optics}} (\bibinfo{publisher}{Peason Education},
  \bibinfo{year}{2008}).

\bibitem[{\citenamefont{Salech and Teich}(2007)}]{SAL07}
\bibinfo{author}{\bibfnamefont{B.~E.~A.} \bibnamefont{Salech}}
  \bibnamefont{and} \bibinfo{author}{\bibfnamefont{M.~C.} \bibnamefont{Teich}},
  \emph{\bibinfo{title}{Fundamentals of Photonics}} (\bibinfo{publisher}{John
  Wiley \& Sons,Inc.}, \bibinfo{year}{2007}).

\end{thebibliography}

\end{document}